\documentclass[aps,twocolumn,nofootinbib,showpacs,prd,aps,10pt]{revtex4}
\usepackage[dvips]{graphicx}
\usepackage[english]{babel}
\usepackage[T1]{fontenc}
\usepackage{mathrsfs}
\usepackage[tbtags]{amsmath}
\usepackage{amssymb}
\usepackage{amsxtra}
\usepackage{amsopn}
\usepackage{latexsym}
\usepackage[mathcal]{eucal}

\newcommand{\BE}{\begin{equation}}
\newcommand{\EE}{\end{equation}}
\newcommand{\BA}{\begin{eqnarray}}
\newcommand{\EA}{\end{eqnarray}}
\newcommand{\Tr}{\mathrm Tr}
\newcommand{\nn}{\nonumber}

\begin{document}

\title{Incorporating fermions in the Gaussian
Effective Potential: the Higgs-Top Sector}

\author{Fabio Siringo}

\affiliation{Dipartimento di Fisica e Astronomia 
dell'Universit\`a di Catania,\\ 
INFN Sezione di Catania,
Via S.Sofia 64, I-95123 Catania, Italy}

\date{\today}
\begin{abstract}
The Higgs-Top model is studied by a non-perturbative variational
extension of the Gaussian Effective Potential that incorporates fermions.
In the limit of a very strong Yukawa coupling the one-loop result is shown
to follow a single-parameter scaling while the gaussian fluctuations give
rise to important deviations from scaling and to a reduction of the vev and
of the top mass. A good general agreement is found with lattice data when
a comparison can be made. The vacuum is shown to be stable for any choice
of the Yukawa coupling, at variance with renormalized perturbation theory.
Analytical results are provided for few observables
like the renormalized mass of the Higgs boson and its wave function renormalization
constant. Extensions to gauge theories like QCD are briefly discussed.
\end{abstract}


\maketitle

\section{introduction}

Above the QCD scale, when all the couplings are small enough,
the Standard Model (SM) of the fundamental interactions can
be studied by perturbation theory.
However, there are energy ranges and sectors of the SM where 
non-perturbative methods are welcome for comparison and
control of the perturbative approximation. For instance
when the QCD becomes strongly coupled, or in the hypothesis of
a large Higgs mass and a strong self coupling of the Higgs sector,
even if almost ruled out by experiments.
An other interesting system is the Higgs-Top sector, where non-perturbative effects could be
found because of the Yukawa interaction $y\approx 0.7$ which is not very small
as for other fermions. 

The most widely used non-perturbative approach is the numerical simulation on a finite lattice, which
allows for an exact treatment of the Lagrangian, but is plagued by many shortcomings like the small
size of the sample, border effects, the lack of any analytical result and some specific problems about incorporating
fermions\cite{genMC2,genMC3,genMC4,genMC5,genMC6, genMC7}. 
For instance a duplication problem does not allow a simple description of a model with
an odd number of fermions or even one only fermion\cite{fermiMC}. Moreover the finite lattice spacing is equivalent to
the existence of a finite energy cut-off that cannot be removed. 

This last point does not seem to be an important 
issue any more, since our modern understanding is that the SM must be considered as an effective model holding up to a 
finite energy scale. The triviality of the scalar theory requires that a finite cut-off should be present in
the model, and even a large cut-off can be simulated on the lattice if the sample size is large enough and
allows for long correlation lengths compared to the lattice spacing. 
On the other hand, the existence of a finite cut-off has opened the way to a class of variational approximations,
since their typical UV problems are cured by the existence of a limited energy range.
The Gaussian Effective Potential (GEP)\cite{schiff,rosen,barnes,GEP} 
is a variational tool that has been recently shown to provide
a very good agreement with lattice data whenever a finite cut-off is used in the analytical derivation\cite{bubble}.
The GEP is not exact, but is based on a non-perturbative approximation, and its predictions are expected
to hold even when perturbation theory breaks down. Moreover the GEP yields simple analytical results and has been recently
shown to sheld some light on the physics of several systems ranging from superconductors\cite{superc1,superc2} 
to magnetic systems\cite{hubbard},
to non-Abelian gauge theories\cite{su2} and the Higgs sector of the SM\cite{var,light,sigma,bubble}.

It would be interesting to extend the GEP to the Higgs-Top sector of the SM, which would require the inclusion
of fermions in the derivation of the effective potential. The problem is of some interest by itself because of
the failure of incorporating fermions that was reported in the past. 
In fact a direct attempt to include fermions was shown to give a result that is equivalent to the perturbative 
one-loop fermionic term of the effective potential\cite{stancu}. 
Thus for fermions the GEP has always been regarded as useless.
Quite recently an hybrid method has been proposed for incorporating fermions in the GEP\cite{hubbard}: 
the technique has been
tested in the framework of the well studied two-dimensional Hubbard Hamiltonian at half filling, describing 
a gas of interacting fermions in the antiferromagnetic phase. The method predicts
the exact magnetization in the strong coupling limit, and for large couplings it improves over other approximations 
like RPA, in close agreement with lattice data. Instead of attempting to evaluate the GEP directly, in this
method the fermions
are integrated exactly, and the resulting effective Lagrangian is expanded in powers of the scalar field.
Then the GEP is evaluated by the usual variational method.
In the strong coupling limit the second order term of the expansion has been shown to be enough for an 
accurate description of the fermionic fluctuations that are included at the Gaussian level\cite{hubbard}.

In this paper the same hybrid technique is used for incorporating fermions in a self-consistent variational
approach to the Higgs-Top model, a toy model containing a self-interacting scalar particle interacting with
a massless fermion through a standard Yukawa interaction. In this model, that mimics the mechanism of
mass generation of the SM, the symmetry breaking is driven by the Yuakawa interaction, and even when the 
classical scalar potential does not show any symmetry breaking, 
the effective potential can have a minimum with a non vanishing 
vacuum expectation value (vev) of the scalar field. Thus while the fermions acquire a mass through the
Yukawa interaction, it is the fermion that drives the symmetry breaking determining the vev. A naive
discussion of symmetry breaking, based on the classical potential, does not work in the present model
where the vev must be determined by the full self-consistent quantum effective potential.

In the next sections the GEP is evaluated for the Higgs-Top model, and the result is compared with 
available lattice simulation data\cite{kuti}
and with the standard one-loop approximation. A full agreement is found with the lattice data when
a comparison can be made. Moreover the method is almost analytical, and explicit analytical results
can be obtained for the pole of the scalar propagator, the wave function renormalization constant
and other relevant observables.
For instance, by inspection of the analytical expression of the effective
potential, the vacuum can be shown to be stable for any choice of the bare
parameters, at variance with renormalized perturbation theory.
The method is non-perturbative, and deviations from the one-loop approximation
can be regarded as a measure of non-perturbative effects that are shown to be
large in the strong coupling limit.

Besides the physical relevance of the present model, the problem of incorporating fermions in a
non-perturbative method is by itself relevant because of the possible extension to non-Abelian
strongly interacting theories like QCD. For instance the GEP has been evaluated for the pure
SU(2) theory\cite{su2} but no previous attempts of incorporating fermions have been reported .
The present method could be explored in that context.

The paper is organized as follows: 
in Section II the method is described in detail;
in Section III the GEP is compared with lattice data; in Section IV the strong coupling limit is
explored and compared with the 1-Loop result; some final remarks and directions for future work
are reported in Section V.  

\section{GEP in the Higgs-Top model}

In the Euclidean formalism the Higgs-Top model is described by the
Lagrangian
\BE
{\cal L}={\cal L}_\phi+{\cal L}_t+{\cal L}_y
\label{L}
\EE
where ${\cal L}_\phi$ is the Lagrangian of a self-interacting scalar
field $\phi$
\BE
{\cal L}_\phi=\frac{1}{2}\partial^\mu\phi\>\partial_\mu\phi+V_c(\phi)
\label{Lphi}
\EE
with a classical potential
\BE
V_c(\phi)=\frac{1}{2}
m_B^2\phi^2+\frac{1}{4!}\lambda_B\phi^4,
\EE

${\cal L}_t$ is the Lagrangian for a set of $N_f$ massless free Fermi
fields $\psi_j$
\BE
{\cal L}_t=-i \sum_{j=1}^{N_f} \bar\psi_j \gamma^\mu \partial_\mu\psi_j
\label{Lt}
\EE
and ${\cal L}_y$ contains a set of Yukawa couplings
\BE
{\cal L}_y=\sum_{j=1}^{N_f} y_j \bar\psi_j \psi_j \phi.
\label{Ly}
\EE
Here we take $y_j=y$ so that the set of Fermi fields $\psi_j$ can be regarded
as a gauge multiplet, and for $N_f=3$ the model describes a quark interacting
with a scalar field. If the symmetry is broken and the scalar field has a
non-vanishing vev, $v=<\phi>$, then the Higgs field $h$ is defined by the shift
$h=\phi-v$ and the fermion acquires the mass $m=yv$. We refer to $m$ as the 
top mass even if the present study applies to any fermion with a Yukawa coupling to
the scalar field.

Let us follow the method of Ref.\cite{hubbard} and integrate  the fermions exactly.
The exact action can be written in terms of a shifted Higgs field
$h=\phi-\varphi$ where $\varphi$ is a generic constant shift.
The field $h$ becomes the standard Higgs field when $\varphi=v$, but we
leave the variable $\varphi$ unconstrained at this stage. Integrating out
the fermions, the bilinear fermionic terms are replaced by an effective action 
 
\BE
S=\int {\cal L} d^4x \to \int {\cal L}_h d^4x+S_{eff}[h]
\label{S}
\EE
where
\BE
{\cal L}_h=\frac{1}{2}\partial^\mu h\>\partial_\mu h+V_c(\varphi+h)
\label{Lh}
\EE
and the effective action $S_{eff}$ is given by
\BE
S_{eff}[h]=-N_f\log \det\left[ G_m^{-1}+yh \right]_{m=y\varphi}
\label{Seff}
\EE
with the fermionic inverse propagator $G^{-1}_m$ defined
as
\BE
G^{-1}_m=-i \gamma^\mu \partial_\mu+m
\label{Gm}
\EE
and the mass set to the value $m=y\varphi$. These equations
are exact and hold for any choice of the variable $\varphi$
as a consequence of the exact integration of the fermionic fields.

It is obvious that the effective action in Eq.(\ref{Seff}) is far too
complicated to be treated exactly. From a formal point of view it
can be expanded in powers of the field $h$

\BE
S_{eff}[h]=-N_f\Tr\log G_m^{-1}
-N_f\sum_{n=1}^{\infty} (-1)^{n+1} y^n \Tr \{(G_m\cdot h)^n\}
\label{expan}
\EE
where the products $G_m\cdot h\cdot G_m \cdot h \cdots$ are non
local products of operators with the internal space variables integrated
over.

The expansion is convergent if the norm $\vert\vert yh G_m \vert\vert< 1$.
We will show that in the strong coupling limit $y\gg \lambda_B\approx 1$,
in units of the cut-off (or of the lattice spacing),
the physical range of parameters, where $m\ll 1$ is reached 
for $0.2 < m_B/y < 0.25 $. 
In fact in that limit the one-loop approximation is entirely characterized by 
the single parameter $m_B/y$ which takes its critical value 
when $m_B/y=\sqrt{N_f}/(2\pi)$, that is $\approx 0.28$ for $N_f=3$.
We will discuss this scaling later, in Section IV.
In that physical range, provided that we do not reach the critical point,
the average $<hh>\approx Z/M_R^2$ where $Z\approx 1$ is a wave function renormalization factor and
$M_R$ is a renormalized mass of order $M_R\approx m_B$, while in units of the cut-off
$<G_m G_m>\approx -1/(4\pi^2)$, so that
\BE
\vert\vert yh G_m \vert\vert^2\approx \frac{y^2}{4}\Tr<h G_m h G_m>\approx
\left(\frac{yZ}{4\pi M_R}\right)^2,
\EE
and for $Z=1$, $M_R\approx m_B$ we see that the expansion makes sense provided that
\BE
\vert\vert yh G_m \vert\vert\approx 0.08\cdot\frac{y}{m_B}<1 
\EE
In the physical
range of parameters where $m_B/y\approx 0.25$ this is never too large, even in the very strong
coupling limit.
 
In fact the expansion in Eq.(\ref{expan}) is not a perturbative expansion in the
parameter $y$, but can be regarded as an expansion in terms of the fluctuating
field $h$. Provided that  $<hh>$ is small, the gaussian fluctuations are enough
and the quartic and higher order terms can be dropped without affecting the 
non perturbative nature of the approximation. This is the case for the
Hubbard model of antiferromagnetism in the strong coupling limit\cite{hubbard}.
The approximation breaks down at the critical point where the fluctuations of
the field $h$ are large and $<hh>$ diverges. However the triviality of the scalar
theory requires that a finite cut-off is retained, that is equivalent to say that
the critical point is never reached. In the physical range of the
parameters, before the critical point is reached,  
even in the strong coupling limit the fluctuations are small and we can neglect
the higher order terms in the expansion. This approximation spoils the variational 
character of the method, but the non-perturbative nature of the approximation is
maintained as for RPA approximation that is very similar
to the present method from a formal point of view.
In principle the approximation can be improved by inclusion of higher order terms in
the expansion, yielding a more complicate set of coupled equations that can be
solved by numerical techniques. However the first non-vanishing contribution comes from the quartic term $<hhhh>$ that is much smaller than the second-order term
in the physical range of the parameters. Its actual value can be evaluated for a more
accurate control of the approximation.

The truncated expansion for $S_{eff}$ reads 

\begin{align}
S^{(2)}_{eff}[h]&=-N_f\Tr\log G_m^{-1}-yN_f\Tr (G_m\cdot h)+   \nonumber\\
&+\frac{1}{2} y^2 N_f\Tr (G_m\cdot h\cdot  G_m\cdot h)\,
\label{Trunc}
\end{align}
and our starting point is the vacuum-to-vacuum amplitude
\BE
Z=\int {\cal D}_h\> e^{\displaystyle{-\left[\int {\cal L}_h d^4x+S^{(2)}_{eff}[h]\right]}}.
\label{z}
\EE
Inserting a gaussian trial functional, the amplitude $Z$ is written as
\BE
Z=\int {\cal D}_h\> e^{\displaystyle{-\left[\frac{1}{2}\int h(x) g^{-1}(x,y) h(y) d^4x d^4y
+S_{int}[h]\right]}}
\label{zint}
\EE
where $g^{-1}(x,y)$ is a trial inverse propagator, and
\begin{align}
S_{int}[h]&=-\frac{1}{2}\int h(x)g^{-1}(x,y)h(y) d^4x d^4y+\nonumber\\
&+\int {\cal L}_h d^4x+S^{(2)}_{eff}[h]\,
\end{align}
that is regarded as an interaction term.
We define the gaussian average of a generic operator ${\cal O}$ according to
\BE
\langle {\cal O} \rangle=\frac{1}{Z_0}\int {\cal D}_h\> {\cal O}
\>e^{\displaystyle{-\frac{1}{2}\int h(x)g^{-1}(x,y)h(y) d^4x d^4y}}
\EE
where 
\BE
Z_0=\int {\cal D}_h
\>e^{\displaystyle{-\frac{1}{2}\int h(x)g^{-1}(x,y)h(y) d^4x d^4y}}.
\EE
It follows that $<h>=0$ and $<h(x)h(y)>=g(x,y)$. Moreover $<\phi>=\varphi$
and the variable $\varphi$ is the expectation value of the scalar field.

By Jensen inequality \cite{ibanez,su2} 
the effective potential $V(\varphi)$ is bounded\cite{GEP,stancu2}  by a gaussian functional
\BE
V(\varphi) \le V_{GEP}[g]
\label{bound}
\EE
where the gaussian functional is defined as
\begin{align}
V_{GEP}[g]&=\langle S_{int}\rangle+\nonumber\\
&-\log\int {\cal D}_h\> 
e^{\displaystyle{-\frac{1}{2}\int h(x)g^{-1}(x,y)h(y) d^4x d^4y}}\,
\end{align}
and is equivalent to the first-order effective potential in presence of the
interaction term $S_{int}$.
The bound in Eq.(\ref{bound}) allows for a variational determination of the
trial propagator $g(x,y)$, and the GEP is defined as the minimum of the functional
\BE 
V_{GEP}(\varphi)\equiv V_{GEP}[g_0]
\EE
with the optimal propagator $g_0$ satisfying the gap
equation
\BE
\left(\frac{\delta V_{GEP}}{\delta g}\right)_{\displaystyle{g_0}}=0
\label{gap}
\EE
This stationary condition must hold at each point $\varphi$ and the
resulting propagator $g_0$ depends on the value of $\varphi$. 

A trivial calculation yields
\BE 
V_{GEP}[g]=V_c(\varphi)+V_{1L}(\varphi)+V_0[g]+V_2[g]+V_F[g].
\label{Gfun}
\EE
The first two terms are the classical potential and the one-loop term, and
do not depend on the trial propagator since
\BE 
V_{1L}(\varphi)=N_f\left(\Tr \log G_m\right)_{m=y\varphi}.
\label{1L}
\EE
The third term has no explicit dependence on $\varphi$
\BE 
V_0[g]=I_1[g]+\frac{\lambda_B}{8}I_0^2[g]
+\frac{1}{2}\int\frac{d^4 p}{(2\pi)^4}\left[\frac{g(p)}{g_B(p)}-1\right].
\EE
Here $g(p)$ is the Fourier transform of $g(x,y)$, the bare propagator
is $g^{-1}_B=m^2_B+p^2$ and the integrals $I_0$, $I_1$
are defined as
\BE
I_1[g]=-\frac{1}{2}\int\frac{d^4 p}{(2\pi)^4}\log g(p)
\EE
\BE
I_0[g]=\int\frac{d^4 p}{(2\pi)^4} g(p).
\EE
All the integrals in the four-dimensional Euclidean space are made finite
by insertion of a cut-off $\Lambda$, and taking $p^2<\Lambda^2$.
The fourth term $V_2[g]$ has a quadratic explicit dependence on $\varphi$
\BE
V_2[g]=\frac{\lambda_B}{4}\varphi^2 I_0[g]
\EE
and the last term $V_F[g]$ follows from the fermionic loop in the average
of the quadratic part of Eq.(\ref{Trunc})
\BE
V_F[g]=\frac{y^2 N_f}{2} \int\frac{d^4 p}{(2\pi)^4} K(p)g(p)
\EE
where the kernel $K(p)$ is the one-loop fermion polarization function
\BE
K(p)=\int\frac{d^4 q}{(2\pi)^4}\Tr\{ G_m(p+q) G_m(q)\}_{m=y\varphi}
\label{K}
\EE
and $G_m(p)$ follows by Fourier transform of Eq.(\ref{Gm})
\BE
G_m(p)=(-\gamma_\mu p^\mu+m)^{-1}=\frac{\gamma_\mu p^\mu+m}{p^2+m^2}.
\label{G}
\EE
Exact and approximate expressions for the kernel $K(p)$ are reported in
appendix A.

According to Eq.(\ref{gap}), the functional derivative of Eq.(\ref{Gfun})
gives a gap equation that reads
\BE
g_0^{-1}=p^2+\Omega^2[g_0]+N_f y^2 K(p)
\label{g0}
\EE 
where the mass functional $\Omega^2[g]$ is defined as
\BE
\Omega^2[g]=m_B^2+\frac{1}{2}\lambda_B\varphi^2+\frac{1}{2}\lambda_B I_0[g]
\label{omega}
\EE
and does not depend on $p$, while $K(p)$ does not depend on $g$.
In the limit $y\to 0$ the optimal trial propagator takes the simple
form $g_0^{-1}=g_\Omega^{-1}=p^2+\Omega^2$ where the mass parameter 
$\Omega^2\equiv\Omega^2[g_0]$ is the self-consistent
solution of Eq.(\ref{omega}). In this limit there is no wave function 
renormalization and we recover the GEP for a scalar theory
\BE
V_{GEP}(\varphi)=V_c(\varphi)+V_0[g_\Omega]+V_2[g_\Omega]
\label{GEPsc}
\EE 
where the last two terms have an implicit dependence on $\varphi$ through $g_\Omega$.
This case has been studied in Ref.\cite{bubble} in some detail.

In general the propagator $g_0$ has a more complicate dependence on $p$ as a consequence
of the Yukawa interaction which adds the last term in Eq.(\ref{g0}).
We can study this dependence in some detail in the limit $p\to 0$. In this limit the kernel
K(p) is an analytic function of $p^2$ and can be expanded in powers yielding
\BE
K(p)\approx a_0(m)+a_1(m)p^2+{\cal O}(p^4)
\EE
where the functions $a_i(m)$ are derived in Appendix A and must be evaluated for $m=y\varphi$.
The optimal propagator in Eq.(\ref{g0}) can then be written in the same limit as
\BE
g_0(p)\approx\frac{Z}{p^2+M^2_R}
\label{g0approx}
\EE
where the wave function renormalization constant reads
\BE
Z^{-1}=1+y^2N_f a_1(m)
\label{Z}
\EE
and the renormalized mass is
\BE
M^2_R=Z\left[\Omega^2+y^2 N_f a_0(m)\right].
\label{MR}
\EE

It is instructive to explore the content of the approximation in terms of graphs.
The solution $g_0$ of the gap equation Eq.(\ref{g0}) can be regarded as the solution 
of the Dyson equation 
\BE
g(p)=g_\Omega(p)+g(p)\cdot [-y^2N_f K(p)]\cdot g_\Omega(p)
\EE
which sums up the whole class of ring-diagrams.
From a formal point of view this kind of approximation is equivalent to RPA, but here the 
self-consistent parameter $\Omega^2=\Omega^2[g_0]$ and the corresponding propagator 
$g^{-1}_\Omega=p^2+\Omega^2$ are functions of $\varphi$ while in RPA they are fixed at their
mean-field value. Thus the dependence of the effective potential on $\varphi$ is
different, as are the predictions for the vev.

Inserting back Eq.(\ref{g0}) in Eq.(\ref{Gfun}) the effective potential takes the simple
form
\BE
V_{GEP}(\varphi)=V_c(\varphi)+V_{1L}(\varphi)+
\left[I_1-\frac{1}{8}\lambda_BI_0^2\right]
\EE
where we recognize the classic term, the one-loop correction and the standard
GEP term of a scalar theory\cite{GEP,bubble} (in brackets). 
However here the integrals $I_0$, $I_1$ are evaluated
for $g=g_0$ which is solution of the interacting gap equation Eq.(\ref{g0}), thus
the last term does depend on the coupling $y$, and gives rise to large 
deviations from the simple one-loop result in the strong-coupling limit.

The numerical evaluation of the effective potential can be obtained more easily by 
a simple integration of the first derivative
\BE
\frac{d V_{GEP}}{d\varphi}=
\frac{\partial V_{GEP}}{\partial \varphi}+
\int \left(\frac{\delta V_{GEP}}{\delta g}\right)
\left(\frac{d g}{d \varphi}\right)
\frac{d^4 p}{(2\pi)^4}
\EE
which by insertion of Eq.(\ref{gap}) is equivalent to the partial derivative
\BE
\frac{d V_{GEP}(\varphi)}{d\varphi}=
\frac{\partial V_{GEP}(\varphi)}{\partial \varphi}
\EE
yielding
\begin{align}
\frac{d V_{GEP}}{d\varphi}&=
\left[m_B^2\varphi+\frac{\lambda_B}{3!}\varphi^3+\frac{d V_{1L}}{d \varphi}\right]+ 
\frac{\lambda_B}{2}I_0\varphi+\nn\\
&+\frac{y^2 N_f}{2}\int \frac{d^4 p}{(2\pi)^4}\frac{d K(p)}{d \varphi} g(p)\,.
\label{dV}
\end{align}
Here the trial propagator must be set to its optimal value $g_0$ for any $\varphi$ point, yielding
an implicit dependence on that variable. 
Explicit expressions for the derivatives of $K(p)$ and $V_{1L}$ are reported in appendix B and C 
respectively.

\begin{figure}[ht]
\includegraphics[height=7cm, width=8cm]{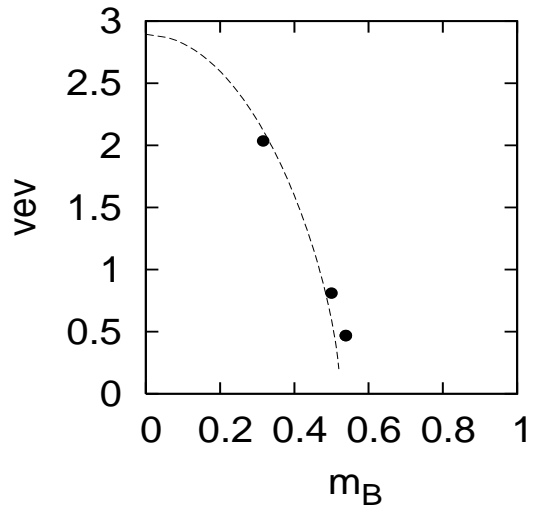}
\caption{\label{Fig1} 
The vev as a function of the bare mass $m_B$ in units of $\Lambda/c$ for 
$\lambda_B=0.1$, $y=0.5$, and for a scale factor $c=2.34$ (dashed line).
With that choice of scale the GEP interpolates the lattice data of Ref.\cite{kuti}
(circles).
}
\end{figure}


\section{Comparison with lattice data}
For any set of lagrangian parameters the GEP can be obtained by solving the gap equation Eq.(\ref{g0})
and integrating the derivative in Eq.(\ref{dV}). In order to compare with lattice data some care 
must be taken in the choice of the energy units. While we would prefer units
of the cut-off $\Lambda$, lattice data are usually reported in units of the lattice spacing $a$. 
Actually the cut-off can be thought as defining an effective lattice spacing $a=c/\Lambda$, with an
unknown constant scale factor $c=a\Lambda$ of order unity that depends on the approximation scheme
and can be determined by a direct comparison
with the lattice data. Once the constant scale factor
is fixed, the predictions of the GEP can be compared with the lattice data\cite{bubble}.
Some lattice data on the Higgs-Top model are discussed in Ref.\cite{kuti}. 
For a comparison we use the same set of parameters: $y=0.5$, $\lambda_B=0.1$ and
$m_B^2=0.1$. As in other lattice calculations, they take a large even number of fermions because of
a well known duplication problem. We use the same value $N_f=8$ in this section.
The scale factor has been determined as $c=2.34$ by a fit of the vev as shown in Fig.1.
Here the vev is defined as the minimum point of the GEP, and is reported in units of $\Lambda/c$ in order
to fit the lattice data.

\begin{figure}[ht]
\includegraphics[height=8cm, width=7cm, angle=-90]{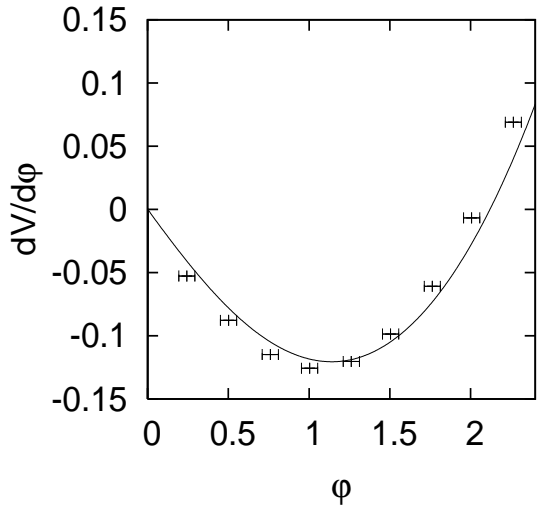}
\caption{\label{Fig2} 
The derivative of the effective potential $dV/d\varphi$ is reported as 
a function of the variable $\varphi$ in units of $\Lambda/c$ with $c=2.34$,
and for the parameter set $\lambda_B=0.1$, $y=0.5$, $m_B^2=0.1$, $N_f=8$.
The GEP (solid line) is compared with the lattice data points of Ref.\cite{kuti}
(error bars).
}
\end{figure}

\begin{figure}[ht]
\includegraphics[height=8cm, width=7cm, angle=-90]{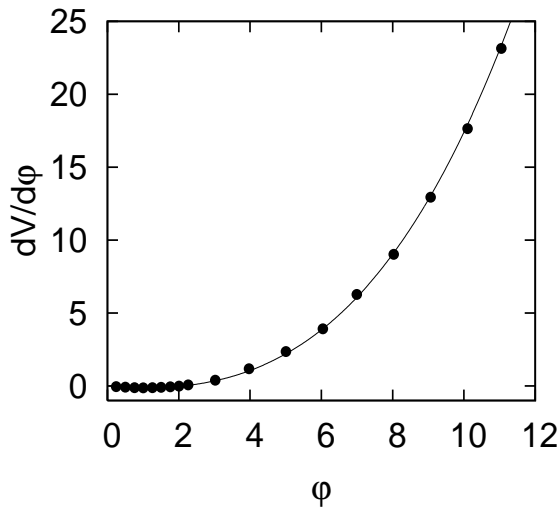}
\caption{\label{Fig3} 
The derivative of the effective potential $dV/d\varphi$ is reported as 
a function of the variable $\varphi$ in units of $\Lambda/c$ with $c=2.34$,
and for the same parameter set as for Fig.2, but for a wide range of $\varphi$.
The GEP (solid line) is compared with the lattice data points of Ref.\cite{kuti}
(circles). The derivative is positive for large values of $\varphi$, even
when $\varphi>\Lambda$, and no vacuum instability occurs in the GEP
}
\end{figure}

\begin{figure}[ht]
\includegraphics[height=8cm, width=7cm, angle=-90]{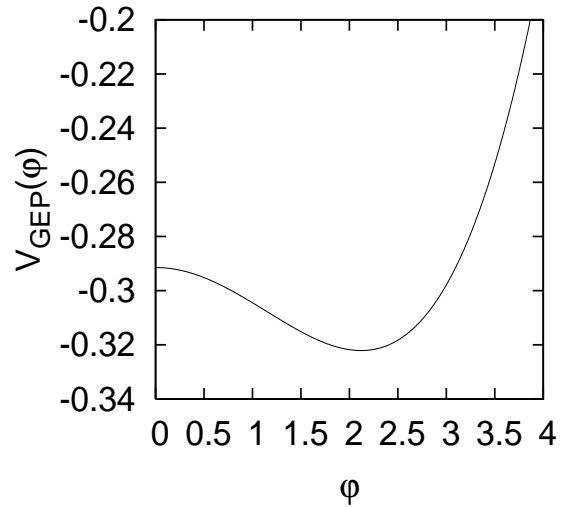}
\caption{\label{Fig4} 
The GEP $V_{GEP}(\varphi)$ is shown as 
a function of the variable $\varphi$ in units of the effective lattice spacing $a=c/\Lambda$ 
with $c=2.34$, and for the same parameter set as for Fig.2.}
\end{figure}

With the scale factor fixed, the derivative of the effective potential is evaluated by
Eq.(\ref{dV}) and reported in Fig.2 for the parameter set $\lambda_B=0.1$, $y=0.5$, $m_B^2=0.1$,
$N_f=8$,
in units of the effective lattice spacing $a=c/\Lambda$.
The lattice data of Ref.\cite{kuti} are shown in the same figure for comparison.
We find a good general agreement, with no need for any other tuning of the parameters.
Unfortunately we only found detailed
lattice data for comparison in this weak-coupling limit
where the variational method does not add to much to the usual perturbative treatment
of the model. 
Some recent lattice data have been reported in the strong coupling limit but
for the special limit cases $\lambda_B\to\infty$ and $\lambda_B= 0$\cite{jansen}.
Strong-coupling deviations from the one-loop predictions will be discussed in the 
next section. 

However, even in this weak coupling limit, at variance with the perturbative
method that predicts an unstable vacuum for large values of $\varphi$, the GEP is perfectly
bounded and predicts a stable vacuum as shown in Fig.3 in agreement with the lattice
calculations, confirming that the instability of the vacuum is not a consequence of the
interaction with the fermions, but a misleading internal problem of the standard perturbative
approximation, emerging when the field is much larger than the cut-off\cite{kuti}.
Actually vacuum instability emerges in the renormalized perturbation theory
because the renormalized couplings are allowed to assume any value, whereas
in bare theories, like the present variational method or lattice calculations, only
a limited range of renormalized couplings are possible, which vanish logarithmically
with the cut-off.
By inspection of Eq.(\ref{dV}) we observe that in the unphysical limit of
$\varphi \gg \Lambda$ the integrals become irrelevant as they are cut at
a relatively small value of $\Lambda$. The negative unstabilizing one-loop term vanishes as $\sim 1/\varphi$ in that limit according to Eq.(\ref{dV1L}), and the
leading term in Eq.(\ref{dV}) is the positive classic $\varphi^3$ term. Thus the effective potential
cannot be unbounded for large values of the field.

For the same set of parameters the GEP is reported in Fig.4 in units of the effective
lattice spacing. The minimum is at $\varphi=v\approx 2.1$ and in the same units the top mass
is $m=yv\approx 1$. It is remarkable that the breaking of symmetry is entirely driven by the
Yukawa interaction with the fermions: here $m_B^2>0$, the classical potential is symmetric and 
would predict a vanishing vev. Thus in the SM the symmetry is broken because of the large Yukawa coupling of 
the top quark $y\approx 0.7$ that gives rise to a finite vev and gives back a mass to all the fermions.
There is no need for unphysical negative bare squared masses $m^2_B<0$ in the Lagrangian, that are quite
hard to be explained for a free field theory. Moreover in several extensions of the SM the Higgs sector
can be quite simplified, because there is no need to assume that the classical potential has a mimimum
at a broken symmetry point $\varphi\not=0$. For instance, the minimal left-right symmetric extension
of the SM\cite{siringoLR1,siringoLR2} does not require the existence of scalar bi-doublet fields, 
and a model with only two doublets
is perfectly viable\cite{siringoLR3,siringoLR4}. 

\section{Strong coupling limit}

The GEP has already been studied for large values of the self-coupling $\lambda_B$ of
the scalar field in the past\cite{bubble}. The method has been shown to be reliable for large
couplings, even if these are almost ruled out by the recent experimental evidence of
a light Higgs mass. In this section we would like to explore the limit of a strong
Yukawa coupling, much larger than $\lambda_B$. We take $\lambda_B=0.1$, and $N_f=3$
in order to represent a QCD quark multiplet like the top quark. Moreover we use units
of $\Lambda$ in this section ($c=1$), and explore the model for a large coupling $y$ and
generic values of the bare mass. Thus in these units we are left 
with two free parameters, $y$ and $m^2_B$. However, the phyisical requirement of a
non vanishing small top mass, quite smaller than the cut-off ($m\ll 1$ in our units), decreases our
degree of freedom, and limits the ratio of the free parameters in the range $0.2<m_B/y<0.25$.
In fact the physical masses can be regarded as the inverse of correlation lengths $\xi=1/m$, and
the condition $m\ll 1$ is equivalent to the requirement that $\xi$ is much larger than the
effective lattice spacing. Only in that limit the model makes sense.
Actually, we can show that the predictions of the one-loop approximation only depend
on the single parameter $y/m_B$ in the strong coupling limit $y\gg \lambda_B\approx 1$.
When the physical top mass $m=yv$ is displayed as a function of $m_B/y$,
the one-loop data collapse on a single curve in the strong coupling limit, and do not
depend on $y$ as dispayed in Fig.5.

\begin{figure}[ht]
\includegraphics[height=8cm, width=7cm, angle=-90]{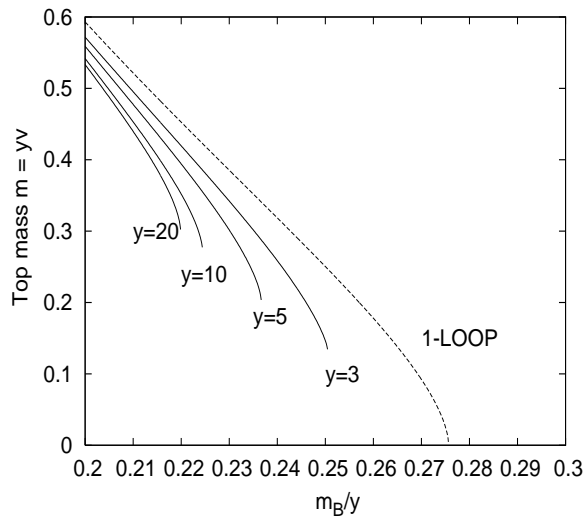}
\caption{\label{Fig5} 
The top mass as 
a function of the parameter $(m_B/y)$, for a range of Yukawa couplings going
from $y=3$ to $y=20$, according to the GEP. 
Energies are in units of the cut-off $\Lambda$. The one-loop data
collapse on a single curve (dashed).}
\end{figure}

Evidence for this strong coupling scaling can be found by a simple
analysis of the single terms contributing to the effective potential.
By taking the mass $m=y\varphi$ as independent variable, the derivative of
the effective potential follows from Eq.(\ref{dV})
\begin{align}
\frac{1}{m}\frac{d V_{GEP}}{d m}&=
\left[
2\left(\frac{d V_{1L}}{d m^2}\right)+
\left(\frac{m_B}{y}\right)^2+
\frac{\lambda_B}{3!y^4}m^3
\right]+\nn\\ 
&+\frac{\lambda_B}{2y}I_0+
y^2 N_f \int \frac{d^4 p}{(2\pi)^4}\frac{d K(p)}{d m^2} g(p)\,.
\label{dVm}
\end{align}
We recognize a first term in square brackets arising from the classical
plus one-loop potential; a second term of order $\sim \lambda_B/y$, arising
from the standard GEP correction to the scalar potential; a third term
of order $\sim y^2$,
arising from our non-perturbative treatment of the Yukawa interaction.
We observe that $V_{1L}$ and $K$ have no explicit dependence on $y$ and $m_B$
when written as functions of the variable $m$.
Thus at one-loop, in the strong coupling limit $y\gg \lambda_B$, the derivative 
of the effective potential reads
\BE
\frac{d V}{d m}=
m\left[
2\left(\frac{d V_{1L}}{d m^2}\right)+
\left(\frac{m_B}{y}\right)^2
\right]
\label{dVm2}
\EE
and only depends on the ratio $m_B/y$.
The stationary points of the potential are the solution of the equation
$dV/dm=0$. When the symmetry is broken we find a maximum at $m=0$ and a minimum
given by the solution of the equation
\BE
\left(\frac{d V_{1L}}{d m^2}\right)=
-\frac{1}{2}\left(\frac{m_B}{y}\right)^2,
\EE
thus a plot of the one-loop top mass $m$ at the minimum, as a function of the single
parameter $(m_B/y)$, must follow a single curve for any strong coupling $y$.

\begin{figure}[ht]
\includegraphics[height=8cm, width=7cm, angle=-90]{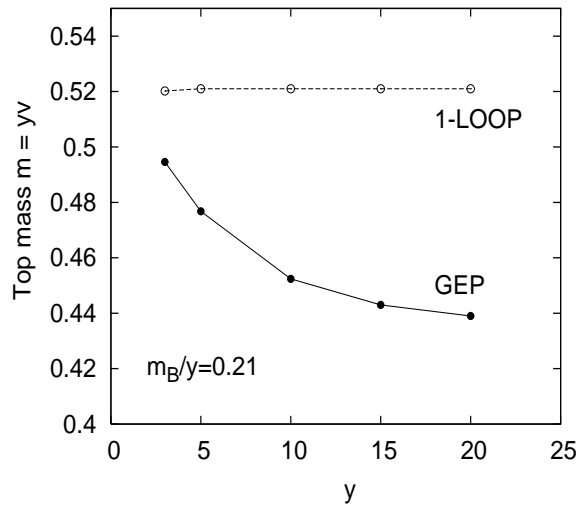}
\caption{\label{Fig6} 
The top mass as 
a function of the Yukawa coupling $y$, for a typical value of the
parameter $(m_B/y)=0.21$. The prediction of the GEP (solid circles) is compared
with the steady one-loop data (open circles). Masses are in units of the cut-off.}
\end{figure}

\begin{figure}[ht]
\includegraphics[height=8cm, width=7cm, angle=-90]{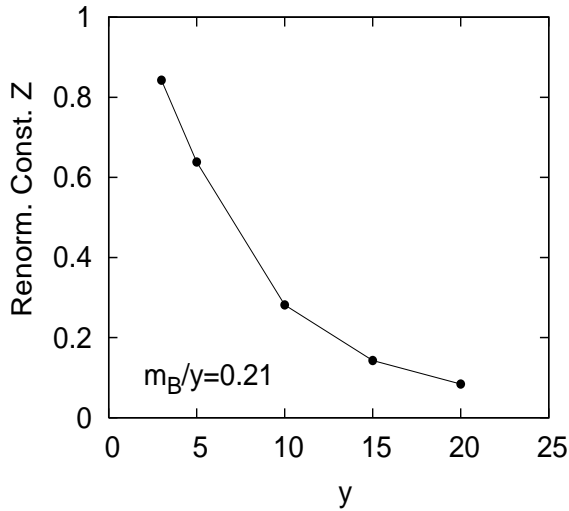}
\caption{\label{Fig7} 
The wave function renormalization constant $Z$ is reported according to Eq.(\ref{Z})
as a function of the Yukawa coupling,  for a typical value of the 
parameter $(m_B/y)=0.21$.}
\end{figure}

This single curve predicts a universal critical point $(m_B/y)_c$ for
the single parameter,
when the maximum and minimum coincide:
\BE
\left(\frac{d V_{1L}}{d m^2}\right)_{m=0}=
-\frac{1}{2}\left(\frac{m_B}{y}\right)^2,
\EE
and from the explicit expression in Eq.(\ref{dV1L})
we obtain
\BE
\left(\frac{m_B}{y}\right)_c=\frac{\sqrt{N_f}}{2\pi}
\EE
that for $N_f=3$ is $(m_B/y)_c\approx 0.276$.
In this strong coupling regime, deviations from the single parameter scaling can only arise 
from the non-perturbative last term in Eq.(\ref{dVm}). Thus deviations from
scaling can be used as a measure of the non-perturbative effects in the GEP.  
The top mass $m$, as emerging from the GEP calculation, 
is reported in Fig.5 as a function of the scaling parameter
$m_B/y$ for several values of the Yukawa coupling ranging from $y=3$ to $y=20$.
All the one-loop data collapse on the dashed line.
Deviations from the scaling are present when the GEP is considered, but they are
small for the phenomenological value $y\approx 0.7$. These deviations cannot be neglected
in the range of large couplings. In Fig.6 the top mass is reported as a function
of the Yukawa coupling $y$ for a typical value of the parameter $m_B/y=0.21$. We notice
the different behaviour of one-loop and GEP approximations: the top mass is reduced
by fluctuations. Moreover the fluctuations give rise to a decreasing of the wave function
renormalization constant $Z$ according to Eq.(\ref{Z}) as shown in Fig.7. 
This effect is negligible for weak
couplings, but becomes very large in the strong coupling limit.

\begin{figure}[ht]
\includegraphics[height=8cm, width=7cm, angle=-90]{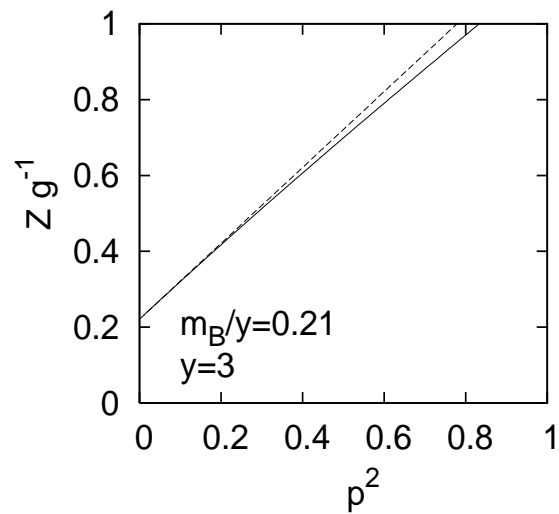}
\caption{\label{Fig8} 
The renormalized inverse propagator $Zg^{-1}$ is shown as a function
of $p^2$, in units of the cut-off $\Lambda$, for a typical value
of the parameter $(m_B/y)=0.21$, and for a moderate coupling $y=3$ (solid line).
Here $g$ is the optimal propagator solving Eq.(\ref{g0}).
The linear approximate inverse propagator of Eq.(\ref{g0approx}) is shown for
comparison (dashed line).}
\end{figure}

\begin{figure}[ht]
\includegraphics[height=8cm, width=7cm, angle=-90]{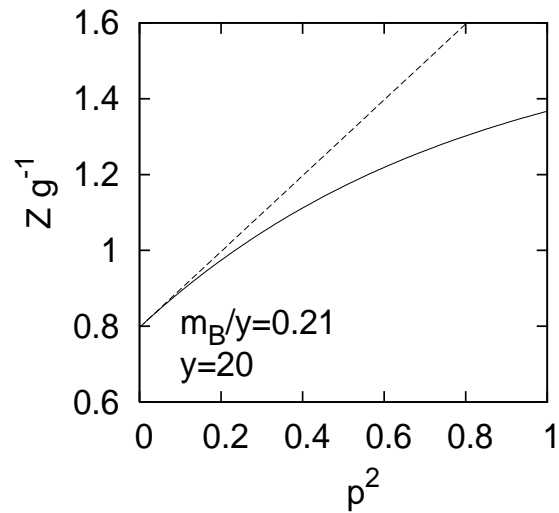}
\caption{\label{Fig9} 
The same as Fig.8 but for a strong Yukawa coupling $y=20$ (solid line).
The linear approximate inverse propagator is shown for
comparison (dashed line).}
\end{figure}

More insights come from a closer study of the Higgs propagator
$g_0(p)$ that solves Eq.(\ref{g0}). The renormalized inverse propagator $Zg_0^{-1}$
is reported as function of $p^2$ for a typical value of the parameter
$m_B/y=0.21$ in Figs.8 and 9. The approximate linear expression in
Eq.(\ref{g0approx}), derived for small values of $p$, is shown for
comparison. In Fig.8, for a moderately large Yukawa coupling $y=3$,
deviations from linearity are small and only 
occur for very large values of $p^2$. Deviations are large in the strong coupling limit,
as shown in Fig.9 for $y=20$. 

A shortcoming of the approximation comes from the break-down of the convergence
criterion when approaching the critical point. Before reaching the critical point
the fluctuations of the Higgs field become large enough to make the expansion
in Eq.(\ref{expan}) useless. In fact in Fig.5 the plots do not reach the critical point,
but we checked that the criterion of convergence is satisfied for the reported data.
From a technical point of view the renormalized mass in Eq.(\ref{MR}) becomes small
when approaching the critical point, and it vanishes before reaching the transition. 
A real pole occurs in the Higgs propagator and the integrals diverge. 
However this is just a 
sign that the Higgs fluctuations are too large and the expansion in Eq.(\ref{expan})
does not hold any more. In our calculation we never let the renormalized mass 
be too small, in
order to fullfill the convergence criterion. Thus our data are still reliable 
even if they cannot span the entire range of the free parameters.

\section{Conclusions}

We have studied the Higgs-Top model by a non-perturbative variational extension
of the GEP that incorporates fermions. While the pure GEP is known to give
trivial results for the fermions\cite{stancu}, some effects of their fluctuations
have been included in the GEP by an hybrid method: fermions are integarted out
exactly and the resulting effective action is expanded in powers of the
fluctuating Higgs field. Even in the strong coupling limit a second order expansion
provides reliable predictions for a large range of
the free parameters. By fixing an effective lattice spacing, the GEP is found in good general agreement with
the available lattice data for the model. 

In the strong coupling limit the gaussian
fluctuations reduce the vev and the top mass, as displayed by a comparison with the one-loop
approximation. In this limit the one-loop data are seen to follow a single-parameter
scaling, while the gaussian fluctuations give rise to important deviations from scaling.

At variance with renormalized perturbation theory,
this study confirms that the Higgs-Top model has a perfectly stable vacuum even in
the strong coupling limit and that, in presence of Yukawa couplings with fermions, there 
is no need for a finite vev in the classical potential.
In fact a non-vanishing vev can be predicted even when $m_B^2>0$ and the classical
potential has no broken symmetry vacuum. Thus fluctuations are very relevant for determining
the correct vev, and naive discussions based on the classical potential cannot be used for
the Higgs-Top model.

On general grounds, inclusion of fermions in a non-perturbative and almost analytical calculation
is of interest by itself. Moreover the same method could be used 
in the framework of effective theories\cite{buccheri} or in
a low-energy study of gauge
theories like QCD when the perturbative approximation breaks down. While the GEP was derived for
the $SU(2)$ gauge theory in the past\cite{su2}, no previous attempt has been reported for
incorporating fermions. Thus the present study could be extended in that direction.

\appendix

\section{Polarization function $K(p)$}
\label{app:K}
The fermionic polarization function $K(p)$ has been calculated exactly, integrating
inside the hyper-sphere $p^2<\Lambda^2$ in the four-dimensional Euclidean space.
Inserting Eq.(\ref{G}) in Eq.(\ref{K}) and evaluating the trace
\BE 
K(p)=4\int_\Lambda\frac{d^4 q}{(2\pi)^4}
\frac{m^2-q^2-q\cdot p}
{(m^2+q^2)(m^2+q^2+p^2+2q\cdot p)}.
\label{K2}
\EE
Here the Feynman trick cannot be used because of the finite cut-off. However by
a tedious but straightforward calculation the exact polarization function can be written
as
\BE
K(p)=K_0+K_1(p)+K_2(p)+K_3(p)
\label{K3}
\EE
where $K_0$ is the constant
\BE
K_0=-\frac{1}{8\pi^2}\left[ \Lambda^2-m^2\log\frac{\Lambda^2+m^2}{m^2} \right]
\EE
and the three functions $K_i(p)$ follow
\BE
K_1(p)=\frac{4m^2+p^2}{16\pi^2}\log\frac{\Lambda^2+m^2}{m^2}+
\frac{\Lambda^2}{32\pi^2p^2}\left(6m^2-\Lambda^2\right)
\EE
\BE
K_2(p)=\frac{\beta\delta+p^4-m^4}{32\pi^2 p^2} + \frac{m^2}{8\pi^2}\log\frac{\beta+\delta}{2m^2}
\EE
\BE
K_3(p)=-\frac{4m^2+p^2}{16\pi^2 p^2} J(p)
\EE
where $\beta$ and $\delta$ are the functions 
\BE
\beta(p)=m^2+\Lambda^2-p^2
\EE
\BE
\delta(p)=\sqrt{\beta^2+4m^2p^2}
\EE
and $J$ is the integral
\BE
J(p)=\int_{m^2}^{m^2+\Lambda^2}\frac{dx}{x} \sqrt{(x-p^2)^2+4m^2p^2}.
\EE
The integral $J$ has been evaluated in Ref.\cite{bubble} and its explicit
expression is
\BE
J(p)=\frac{\beta+\delta-2m^2}{2}-p^2\rho
\EE
where $\rho$ is the function
\begin{align}
\rho(p)&=\log\left\vert\frac{t_2}{t_1}\right\vert-\frac{m}{p}\left(\frac{1}{t_2}-\frac{1}{t_1}\right)+
\nonumber\\
&+\sqrt{1+\frac{4 m^2}{p^2}} \log\left\vert
\frac{(t_2-t_-)(t_1-t_+)}{(t_2-t_+)(t_1-t_-)}\right\vert \,
\end{align}
and its arguments are
\BE
t_\pm=-\frac{p}{2m}\pm\sqrt{1+\frac{p^2}{4m^2}}
\EE
\BE
t_1=\frac{m}{p}\>;\qquad t_2=\frac{\beta+\delta}{2mp}.
\EE

It can be easily seen that the singular terms cancel in Eq.(\ref{K3})
and the resulting function $K(p)$ is an analytic function of $p^2$. Its
expansion in powers of $p^2$ reads
\BE
K(p)=\sum_{n=0}^{\infty} a_n(m) p^{2n}
\EE
where the coefficients $a_n(m)$ are functions of $m^2$. In units of the cut-off
$\Lambda$ the first two coefficients of the expansion are
\BE
a_0(m)=-\frac{1}{4\pi^2}\left[1+\frac{2m^2}{1+m^2}-3m^2\log\frac{1+m^2}{m^2}\right]
\EE
\BE
a_1(m)=\frac{1}{8\pi^2}\log\frac{1+m^2}{m^2}-\frac{(6m^4+21m^2+7)}{48\pi^2(1+m^2)^3}.
\EE


\section{Derivative of $K(p)$}
\label{app:dK}

Explicit expressions for the exact and approximate polarization function
$K(p)$ have been reported in Appendix A. The derivative can be written as
\BE 
\frac{d K(p)}{d \varphi}=2ym\frac{dK(p)}{d m^2}
\EE
where
\BE
\frac{ d K(p)}{d m^2}=\frac{ d K_0}{d m^2}+
\frac{ d K_1(p)}{d m^2}+
\frac{ d K_2(p)}{d m^2}+\frac{ d K_3(p)}{d m^2}
\label{dK}
\EE
and the single terms $K_i$ are given in Appendix A.
By an explicit calculation we obtain
\BE
\frac{ d K_0}{d m^2}=-\frac{1}{8\pi^2}\left[ \frac{\Lambda^2}{\Lambda^2+m^2}
-\log\frac{\Lambda^2+m^2}{m^2} \right],
\EE

\begin{align}
\frac{ d K_1(p)}{d m^2}&=\frac{1}{4\pi^2}\left[\log\frac{\Lambda^2+m^2}{m^2}-
\frac{\Lambda^2}{\Lambda^2+m^2}\right]+\nn \\
&-\frac{\Lambda^2 p^2}{16\pi^2m^2(\Lambda^2+m^2)}+
\frac{3 \Lambda^2}{16\pi^2 p^2}\,,
\end{align}

\begin{align}
\frac{ d K_2(p)}{d m^2}&=-\frac{1}{8\pi^2}-\frac{m^2}{16\pi^2 p^2}+
\frac{(\Lambda^2+m^2)^2}{16\pi^2 p^2 \delta}+ \nn \\
&+ \frac{m^2-\Lambda^2}{16 \pi^2 \delta}
+\frac{1}{8\pi^2}\log\frac{\beta+\delta}{2m^2}+\nn \\
&+\frac{m^2}{8\pi^2\delta}\left[1+\frac{2p^2}{\beta+\delta}\right]\,,
\end{align}

\BE
\frac{ d K_3(p)}{d m^2}=
-\frac{(4m^2+p^2)}{16\pi^2 p^2}\frac{ d J(p)}{d m^2}
-\frac{J(p)}{4 \pi^2 p^2},
\EE
where the functions $\beta$, $\delta$ and $J$ are defined in Appendix A.
The derivative of $J$ follows as

\begin{align}
\frac{ d J(p)}{d m^2}&=
\frac{\beta+2p^2-\delta}{2\delta}-\frac{2p^2(\Lambda^2+2m^2)}{m^2(\Lambda^2+m^2)}+\nn \\
&+p^2\frac{\beta+2m^2+\delta}{\beta+\delta}\left[\frac{1}{m^2}
-\frac{\beta+2p^2+\delta}{\delta(\beta+\delta)}\right]+\nn \\
&-\frac{2p}{\sqrt{p^2+4m^2}} \log\left\vert
\frac{(t_2-t_-)(t_1-t_+)}{(t_2-t_+)(t_1-t_-)}\right\vert+\nn \\
&+\frac{p^2\left[(\beta+2p^2)(4m^2+p^2)+(4m^2-p^2)\delta\right]}
{(m^2+\Lambda^2)(\beta+\delta)\delta}\,,
\end{align}
where the functions $t_\pm$, $t_1$, $t_2$ are defined in Appendix A.

\section{One-loop potential $V_{1L}$ and its derivative}
\label{app:V1L}

The one-loop term $V_{1L}(\varphi)$ is defined in Eq.(\ref{1L}) and its explicit
evaluation, in terms of $m=y\varphi$, yields the integral
\BE
V_{1L}(m)=-2N_f\int_0^\Lambda\frac{2\pi^2p^3dp}{(2\pi)^4}\log (m^2+p^2)
\EE
and in units of the cut-off $\Lambda$
\begin{align}
V_{1L}(m)&=\frac{N_f}{32\pi^2}\left[1-2\log(1+m^2)\right]+\nn \\
&-\frac{N_f m^2}{16\pi^2}
+\frac{N_f m^4}{16\pi^2}\log\frac{1+m^2}{m^2} \,.
\end{align}

The derivative follows
\BE
\frac{d V_{1L}}{d m^2}=\frac{N_f m^2}{8\pi^2}\log\frac{1+m^2}{m^2}-
\frac{N_f}{8\pi^2}
\label{dV1L}
\EE
so that
\BE
\left(\frac{d V_{1L}}{d m^2}\right)_{m=0}=-\frac{N_f}{8\pi^2}.
\EE


\begin{thebibliography} {99}


\bibitem{genMC2} R. Gupta, arXiv:hep-lat/9807028.
\bibitem{genMC3} D. N. Petcher, arXiv:hep-lat/9301015v1.
\bibitem{genMC4} U.-J. Wiese, in Foundations and New Methods in Theoretical Physics, Saalburg, (2009).
\bibitem{genMC5} R. V. Gavai, Pramana {\bf 61}, 889 (2003).
\bibitem{genMC6} R. V. Gavai, Pramana {\bf 67}, 885 (2006).
\bibitem{genMC7} T. Onogi, Int. J. Mod. Phys. A {\bf 24}, 4607 (2009).
\bibitem{fermiMC} S. Chandrasekharan and U.-J. Wiese,  Prog. Part. Nucl. Phys. {\bf 53}, 373 (2004),
arXiv:hep-lat/0405024v1.
\bibitem{schiff} L.I. Schiff, Phys. Rev. {\bf 130}, 458 (1963).
\bibitem{rosen}G. Rosen, Phys. Rev. {\bf 172}, 1632 (1968).
\bibitem{barnes}T. Barnes and G. I. Ghandour, Phys. Rev. D {\bf  22 }, 924 (1980).
\bibitem{GEP} P.M. Stevenson, Phys. Rev. D {\bf 32}, 1389 (1985).
\bibitem{bubble} F. Siringo and L. Marotta, Int. J. Mod. Phys. {\bf A25}, 5865 (2010), arXiv:0901.2418v2.
 \bibitem{superc1} M. Camarda, G.G.N. Angilella, R. Pucci, F. Siringo, 
Eur. Phys. J. B {\bf 33}, 273 (2003).
 \bibitem{superc2} L. Marotta, M. Camarda, G.G.N. Angilella and F. Siringo, 
Phys. Rev. B {\bf 73}, 104517 (2006).
\bibitem{hubbard} L. Marotta and F. Siringo,Mod. Phys. Lett. B, {\bf 26}, 1250130 (2012), arXiv:0806.4569v3.
\bibitem{su2} F. Siringo, L. Marotta, Phys. Rev. D {\bf 78}, 016003 (2008).
\bibitem{var} F. Siringo,  Phys. Rev. D {\bf 62}, 116009 (2000).
\bibitem{light} F. Siringo, Europhys. Lett. {\bf 59}, 820 (2002).
\bibitem{sigma} L. Marotta and F. Siringo, Eur. Phys. J. C {\bf 44} 293, (2005).
\bibitem{stancu} I. Stancu, Phys. Rev. D {\bf 43}, 1283 (1991).
\bibitem{kuti} Z. Fodor, K. Holland, J. Kuti, D. Nogradi, C. Schroeder, arXiv:0710.3151v1. 
\bibitem{ibanez} R. Iba\~nez-Meier, I. Stancu, P.M. Stevenson, Z. Phys. C {\bf 70}, 307 (1996).
\bibitem{stancu2} I. Stancu and P. M. Stevenson, Phys. Rev. D {\bf 42}, 2710 (1990).
\bibitem{jansen} J. Bulava, P. Gerhold, G. W.-S. Hou, K. Jansen, B. Knippschild, C.-J. David Lin, K.-I. Nagai, A. Nagy, K. Ogawa, B. Smigielski, PoS Lattice2011:075 (2011),
arXiv:1111.4544.
\bibitem{siringoLR1} F. Siringo, Eur. Phys. J. C {\bf 32}, 555 (2004); hep-ph/0307320.
\bibitem{siringoLR2} F. Siringo, Phys. Rev. Lett. {\bf 92}, 119101 (2004).
\bibitem{siringoLR3} F. Siringo and L. Marotta, Phys. Rev. D {\bf 74}, 115001 (2006).
\bibitem{siringoLR4} F. Siringo, Particles and Nuclei, Letters , in press,  arXiv:1208.3599.
\bibitem{buccheri} F. Siringo and G. Buccheri, Phys. Rev. D {\bf 86}, 053013 (2012),  arXiv:1207.1906. 

\end{thebibliography}


\end{document}